\documentclass[longnamesfirst,apjl]{emulateapj}
\usepackage{apjfonts,natbib,epsfig}
\tighten

\newcommand{\beq}{\begin{equation}}
\newcommand{\eeq}{\end{equation}}
\newcommand{\bea}{\begin{eqnarray}}
\newcommand{\eea}{\end{eqnarray}}

\begin{document}
\title{Contribution of Stellar Tidal Disruptions to the X-Ray Luminosity Function of Active Galaxies}
\author{Milo\v s Milosavljevi\'c\altaffilmark{1,2}, David Merritt\altaffilmark{3}, and Luis C.~Ho\altaffilmark{4} }
\altaffiltext{1}{Theoretical Astrophysics, Mail Code 130-33, California Institute of Technology, 1200 East California Boulevard, Pasadena, CA 91125.  }
\altaffiltext{2}{Hubble Fellow.}
\altaffiltext{3}{Department of Physics, Rochester Institute of Technology, 84 Lomb Memorial Drive, Rochester, NY 14623.}
\altaffiltext{4}{Observatories of the Carnegie Institution of Washington, 813 Santa Barbara Street, Pasadena, CA 91101.}
\righthead{AGN LUMINOSITY FUNCTION}
\lefthead{MILOSAVLJEVI\'C ET AL.}

\begin{abstract}

The luminosity function of active galactic nuclei has been measured down to luminosities $\sim 10^{42}\textrm{ ergs s}^{-1}$ in the soft and hard X-rays.  Some fraction of this activity is associated with the accretion of the material liberated by the tidal disruption of stars by massive black holes.  We estimate the contribution to the X-ray luminosity function from the tidal disruption process.  While the contribution depends on a number poorly known parameters, it appears that it can account for the majority of X-ray selected AGN with soft or hard X-ray luminosities $\lesssim 10^{43}-10^{44}\textrm{ ergs s}^{-1}$.  If this is correct, a picture emerges in which a significant portion of the X-ray luminosity function of AGN is comprised of sources powered by tidal-disruption at the faint end, while the sources at the bright end are powered by non-stellar accretion.  Black holes with masses $\lesssim 2\times10^6 M_\odot$ could have acquired most of their present mass by an accretion of tidal debris.  In view of the considerable theoretical uncertainty concerning the detailed shape of the light curves of tidal disruption events, we focus on power-law luminosity decay (as identified in candidate tidal disruption events), but we also discuss constant accretion rate models.

\keywords{ accretion --- black holes --- cosmology: observations --- galaxies: active --- galaxies: nuclei }

\end{abstract}

\section{Introduction}
\label{sec:introduction}

\setcounter{footnote}{0}

A main-sequence or giant star passing near a massive black hole (MBH) of mass $M_{\rm bh}\lesssim 10^8M_\odot$ is disrupted by the tidal forces \citep{Hills:75}.  The disruption and the subsequent accretion of stellar debris by the black hole are potential powerful sources of X-ray emission (e.g., \citealt{Meszaros:77,Young:77,Rees:88,Rees:90}).    The emission can be produced in various phases of tidal disruption, including the initial compression of the star \citep{Kobayashi:04}, the interaction of ejected debris with the ambient medium \citep{Khokhlov:96}, in tidal stream collisions and debris fallback onto a nascent accretion disk \citep{Kochanek:94,Lee:96,Kim:99,Li:02}, and in the accretion of disk material onto the black hole \citep{Cannizzo:90,Ulmer:99}.  The purpose of the present study is to estimate the contribution of activity associated with tidal disruptions to the luminosities of typical active galaxies.  

Recently, surveys with {\it HEAO 1}, {\it ASCA}, {\it Chandra}, and {\it XMM-Newton} have yielded luminosity functions (LFs) of active galaxies at low and high redshift in the hard (e.g., \citealt{Ueda:03,Barger:05b}) and soft (e.g., \citealt{Hasinger:05}) X-rays.  The measured X-ray LFs extend to lower luminosities than their optical counterparts because contamination with host galaxy light makes optical detection of low-luminosity AGN difficult.  The LFs are approximately  broken power laws $dN/d\ln L_{\rm X}\propto L_{\rm X}^\mu(L_{\rm X}+L_{{\rm X}\star})^{\nu-\mu} $ with $\nu\approx -2$ and $-1\lesssim \mu \lesssim 0$ and $L_{{\rm X}\star}\sim 10^{43}-10^{44}\textrm{ ergs s}^{-1}$.  The power-law behavior at the bright end has been attributed to the intrinsic structure of quasar luminosity histories.  The luminosity histories (light curves) are convolved with the black hole mass function to calculate the LF (e.g.,~\citealt{Wyithe:03,Hopkins:05}).  The origin of the LF at the faint end, however, remains unknown.

At least a fraction of low-luminosity AGN must be powered by stellar tidal disruption. Here we estimate the contribution of these sources to the X-ray LF of AGN.  This estimate  is an essential step to understanding the origin of activity in AGN with black hole masses $\lesssim 10^7M_\odot$ and ultimately to understanding the origin of low-mass MBHs.

Theoretical investigations of the stellar tidal disruption process, and the detection X-ray flares that are candidate tidal disruption events (e.g., \citealt{Komossa:99,Komossa:04b,Halpern:04}), suggest that the nuclei in which stellar tidal disruption has occurred remain luminous for $1-10\textrm{ yrs}$ after disruption or longer.  The integrated luminosity of the AGN depends on the mass of the stellar debris that remains bound to the black hole, and on the fraction of this mass that ends up accreting onto the black hole in a radiatively-efficient fashion. The bound mass can be a fraction $f\sim 0.25-5$ of the initial stellar mass \citep{Ayal:00}. 

The bound mass inferred in the giant X-ray flare in NGC 5905 using a black-body fallback model is much smaller than the solar mass \citep{Li:02}; in two other candidate flares, RX J1624.9$+$7554 and RX J1242.6$-$1119A, the light curves are consistent with masses $\sim M_\odot$.  \citet{Donley:02} constrained the rate of outbursts by comparing the ROSAT All-Sky Survey with archival pointed observations.  They inferred a rate of $\sim 10^{-5}\textrm{ yr}^{-1}$ per galaxy.
%, and argued that the rate was consistent with estimates for the tidal disruption rate.  
We adopt a theoretical estimate for the tidal disruption rate that is significantly higher (eq.~\ref{eq:n_dot}).

The outline of our calculation of the AGN LF is as follows.  In \S~\ref{sec:disruption_rate}, we estimate the mass function of MBHs and the rate of tidal disruptions.  In \S~\ref{sec:light_curve}, we present simple models for the light curves of tidally disrupted stars.  In \S~\ref{sec:integration}, we calculate the LF and explore its dependence on various parameters in the model.  In \S~\ref{sec:discussion}, we discuss implications of our estimate for the understanding of low-luminosity, X-ray selected AGN.  

\section{Calculation of the Luminosity Function}

\subsection{Tidal Disruption Rate}
\label{sec:disruption_rate}

Stars passing within distance $(\eta^2 M_{\rm bh}/m_\star)^{1/3} R_\star$ of the black hole, where $m_\star$ is the mass of the star, $R_\star$ is its radius, and $\eta\sim 1$ is a numerical factor, are tidally disrupted. Here we estimate the tidal disruption rate in a galaxy with black hole of mass $M_{\rm bh}$ and stellar velocity dispersion $\sigma$.  The disruption rate  depends on the detailed structure of the stellar nucleus of the galaxy.
%\footnote{The tidal disruption rate in nuclei that contain or have contained a binary MBH can be significantly reduced \citep{Merritt:05}.} 
The galaxies relevant to this study are those with $M_{\rm bh}\lesssim M_{\rm max}\equiv 10^8 M_\odot (R_\star/R_\odot)^{3/2}(m_\star/M_\odot)^{-1/2}$ which can disrupt stars.  

In spheroids as faint as that of the Milky Way, the relaxation time within the black hole's radius of dynamical influence $r_{\rm bh}\sim GM_{\rm bh}/\sigma^2$ is short enough that a collisionally-relaxed distribution, $\rho\propto r^{-\gamma}$, where $\gamma\sim 1.5-1.75$, is set up \citep{Bahcall:76,Bahcall:77,MerrittSzell:05}.  This is consistent with what is seen at the Galactic Center \citep{Genzel:03,Schoedel:06}.  But just beyond $r_{\rm bh}$, the Galactic Center density profile steepens to $\rho\propto r^{-2}$, and this is also the slope observed with the {\it Hubble Space Telescope} near the centers of all but the most luminous galaxies \citep{Gebhardt:96,Ferrarese:06}. Furthermore, nuclear relaxation times in galaxies with $M_{\rm bh}\gtrsim 10^7 M_\odot$ generally exceed a Hubble time \citep{Faber:97} and so a Bahcall-Wolf cusp would not form.

The rate of stellar disruptions in stellar density cusps has been studied at various levels  of detail (e.g., \citealt{Frank:76,Lightman:77,Cohn:78,Magorrian:99}).  We adopt the most recent estimate of the  rate in a $\rho\propto r^{-2}$ cusp \citep{Wang:04}
\bea
\label{eq:n_dot}
\Gamma(M_{\rm bh})&\approx& 7\times10^{-4}\textrm{ yr}^{-1} \left(\frac{\sigma}{70\textrm{ km s}^{-1}}\right)^{7/2}\left(\frac{M_{\rm bh}}{10^6 M_\odot}\right)^{-1}\nonumber\\
& &\times \left(\frac{m_\star}{M_\odot}\right)^{-1/3} \left(\frac{R_\star}{R_\odot}\right)^{1/4} ,
\eea
where $m_\star$ and $R_\star$ are the mass and radius of the tidally disrupted stars.
%\footnote{If the two-body collisional diffusion rate of stars with mass $m_\star$ is dominated by stars of different mass, the disruption rate in equation (\ref{eq:n_dot}) is not strictly correct.  We ignore this complication.} 

The disruption rate in equation (\ref{eq:n_dot}) must be weighted by the stellar mass function in the spheroid at distances from the black hole corresponding to the initial pericenter radii of tidally disrupted stars.  These radii are typically similar to $r_{\rm bh}$ \citep{Magorrian:99}.   If the stellar mass function resembles the ``universal'' initial mass function \citep{Kroupa:01}, the disruption is dominated by $m_\star \sim 0.1M_\odot$ stars. However, the mass function near the black hole may differ from the universal initial mass function because higher-mass stars dynamically segregate closer to MBH, and because stars with an unusual initial mass function may form in situ near the MBH (e.g., \citealt[and references therein]{Milosavljevic:04}). 

Mounting observational evidence indicates that the black hole mass and the velocity dispersion of the host stellar spheroid are tightly correlated with best-fit relation $M_{\rm bh}\approx 1.7\times10^8 M_\odot (\sigma/200\textrm{ km s}^{-1})^{4.9}$ (e.g., \citealt[and references therein]{Ferrarese:05}).  We explore departures from this fiducial relation by parameterizing the logarithmic slope of the relation $M_{\rm bh}\propto \sigma^a$.

No direct constraints on the cosmic density of MBH with masses $\lesssim 10^7 M_\odot$ exist. The majority of these MBH should be in spiral galaxies.  The density can be estimated by assuming a relation between the bulge luminosity and the black hole mass of the form $L_{\rm bulge}=AM_{\rm bh}^k$.  Following \citet{Ferrarese:02}, we substitute this in the Schechter LF $\Phi(L)dL=\Phi_0 (L/L_\star)^\alpha e^{-L/L_\star} dL/L_\star$ to obtain a cosmic mass function of black holes
\beq
\label{eq:mbh_mass_function}
\Psi(M_{\rm bh}) dM_{\rm bh} = \Psi_0 \left(\frac{M_{\rm bh}}{M_\star}\right)^{k(\alpha+1)-1} e^{-(M_{\rm bh}/M_\star)^k} \frac{dM_{\rm bh}}{M_\star}
\eeq
where $\Psi_0=k\Phi_0$, $M_\star=(1.27\ \beta L_\star /A)^{1/k}$, and $\beta\equiv L_{\rm bulge}/L\sim 0.3$ for the Hubble type Sab  \citep{Simien:86}  which we use as reference.  The latter is justified because spiral galaxies dominate galaxy LF below $\sim L_\star$ (e.g., \citealt{Nakamura:03}).  For the relation between the blue luminosity of the galaxy and the black hole mass we adopt $L=3900 L_\odot (M_{\rm bh}/M_\odot)^{0.79}$ \citep{Marconi:03}.  
%which in the range $M_{\rm bh}=(10^5-10^8)M_\odot$ is close a similar relation $L=1,300 L_\odot (M_{\rm bh}/M_\odot)^{0.84}$ used in \citet{Tamura:06} to obtain a similar estimate of the black hole mass function.

The number density of $\sim 10^8M_\odot$ black holes estimated using equation (\ref{eq:mbh_mass_function}) agrees with a similar estimate in \citet{Marconi:04}, while the number density of $\sim 10^6M_\odot$ black holes exceeds their estimate by a factor $\sim 3-4$.  This may be a consequence of our using the $B$-band galaxy LF as reference, whereas \citet{Marconi:04} use the $K$-band LF that is flatter at the faint end. However our adoption of a steeper $M_{\rm bh}$--$\sigma$ relation than they do has an opposite effect.

\subsection{The Light Curve of a Tidal Disruption Event}
\label{sec:light_curve}

The light curves of candidate tidal disruption events \citep{Komossa:04b} are characterized by rapid fall in X-ray luminosity on time scales of months to a year, followed by a gradual continued fading over the following decade.  Three of the four candidate events (NGC 5905, RX J1420$+$53, RX J1242$-$11) are consistent with luminosity decay $L_{\rm fb}\propto t^{-5/3}$ expected if the X-ray emission is produced during the fallback of stellar debris onto a nascent accretion disk \citep{Phinney:89,Evans:89}.  The fourth event RX J1624$+$75 exhibits what appears to be faster initial decay followed by very slow fading.  In all cases, the total drop in X-ray luminosity is by a factor $\sim 10^2-10^3$ over about a decade.

The above candidate tidal disruption events have been selected for characteristic, ``outburst''-like light curves (fast rise and slow decay). However the true luminosity evolution of  tidal disruption events is unknown. If the emission during the fallback stage is absorbed or otherwise suppressed, the X-ray light curve will be dominated by thin disk accretion and a quasi-steady flux over a longer period will be expected.  Such events would remain undetected in existing searches that target outburst activity.  To maintain generality, therefore, we consider fallback-like power-law light curves, and also discuss constant accretion rate models. 

The power irradiated during fallback was derived in \citet{Li:02}
\bea
\label{eq:light_curve}
L_{\rm bol}&\approx& 4\times10^{42}\textrm{ ergs s}^{-1} \left(\frac{f}{0.25}\right)\left(\frac{M_{\rm bh}}{10^6M_\odot}\right)\nonumber\\& &\times\left(\frac{m_\star}{M_\odot}\right)^{2/3}
\left(\frac{t}{1\textrm{ yr}}\right)^{-5/3} ,
\eea
where the tidal disruption occurs at $t=0$, and the luminosity reaches peak value and starts decaying as in equation (\ref{eq:light_curve}) at time $t_{\rm peak}\approx 0.02 (M_{\rm bh}/10^6M_\odot)^{1/2}(m_\star/M_\odot)^{-1}(R_\star/R_\odot)^{3/2}\textrm{ yr}$ after tidal disruption.

Although equation (\ref{eq:light_curve}) was derived for the fallback process, we interpret it as a representative power-law decay model for the luminosity evolution, independent of the emission mechanism  To test the sensitivity to the power law index of luminosity decay, we consider decay of the form $L_{\rm bol}\propto t^{-\lambda}$, where $\lambda>1$ is a constant.  We also test the sensitivity to the value of $t_{\rm peak}$ by scaling it via $t_{\rm peak}(\xi)=\xi t_{\rm peak}(0)$.  In both cases we keep the total energy irradiated between $t=t_{\rm peak}$ and $t=\infty$ constant and equal to $f m_\star c^2$.

We also discuss an alternative to power-law light curves.  Accurate theoretical modeling of a time-dependent thin disk emission is difficult. We therefore adopt a crude quasi-steady accretion scenario, in which we assume that accretion at constant rate proceeds ``while the supplies last,'' i.e., until all the bound debris has been accreted.  This is inevitably an oversimplification. Taking into account the redistribution of mass in the disk, \citet{Cannizzo:90} estimate that under certain assumptions  the light curve will be $L_{\rm disk}\propto t^{-1.2}$.  On the other hand, if outer parts of the disk become thermally unstable and neutral, the accretion may be delayed \citep{Menou:01}.  

In the ``while the supplies last'' constant-accretion scenario, we assume that the luminosity is a fraction $\ell$ of the Eddington luminosity $L_{\rm Edd}(M_{\rm bh})$, and that the luminosity is related to the mass accretion rate via $L=\epsilon \dot M c^2$, where as usual $\epsilon$ parameterizes radiative efficiency.  The duration of the active phase is then $\Delta t_{\rm active}= \epsilon f m_\star c^2/\ell L_{\rm Edd}$, implying a model luminosity $L_{\rm bol}(M_{\rm bh},t)=\ell L_{\rm Edd} H(\Delta t_{\rm active}-t)$, where $H(x)$ is the Heaviside step function.

\subsection{Integration over the Black Hole Mass Function}
\label{sec:integration}

The LF of tidal disruption events is obtained by weighting the disruption probability modeled as a Poisson process by the black hole mass function.  In the fallback model, the LF reads
\bea
\label{eq:psi}
\Psi(L_{\rm X})&=&  \int_{M_{\rm min}}^{M_{\rm max}(m_\star) }dM_{\rm bh} 
\int_{t_{\rm peak}(M_{\rm bh})}^{\infty} 
dt\ \Psi(M_{\rm bh})\  \Gamma(M_{\rm bh}) e^{-\Gamma(M_{\rm bh})t} 
\nonumber\\& & \times
\delta[\omega^{-1} L_{\rm bol}(M_{\rm bh},t)-L_{\rm X}] ,
\eea
where $\omega$ is the bolometric correction, $\delta(x)$ is the Dirac $\delta$-function, and the dependence on $f$ and $m_\star$ is implicitly assumed; for the radius of subsolar stars we adopt the relation $R_\star = R_\odot (m_\star/M_\odot)^{0.8}$ (e.g., \citealt{Kippenhahn:90}).  Here, $M_{\rm min}$ is the minimum MBH mass, and as before $M_{\rm max} \propto m_\star^{0.7}$ is the maximum MBH mass near which a star can be disrupted.  We take $M_{\rm min}=10^4 M_\odot$, as systems in which an even smaller black hole is expected have been found to lack one (e.g., \citealt{Merritt:01,Valluri:05}).  For the bolometric correction in the $0.5-2\textrm{ keV}$ band, we take $\omega=45$, which at low luminosities is approximately independent of luminosity \citep[and references therein]{Marconi:04}, although in reality the bolometric correction could be correlated with $L_{\rm bol}/L_{\rm Edd}$ rather than $L_{\rm bol}$.

\begin{figure*}
\plotone{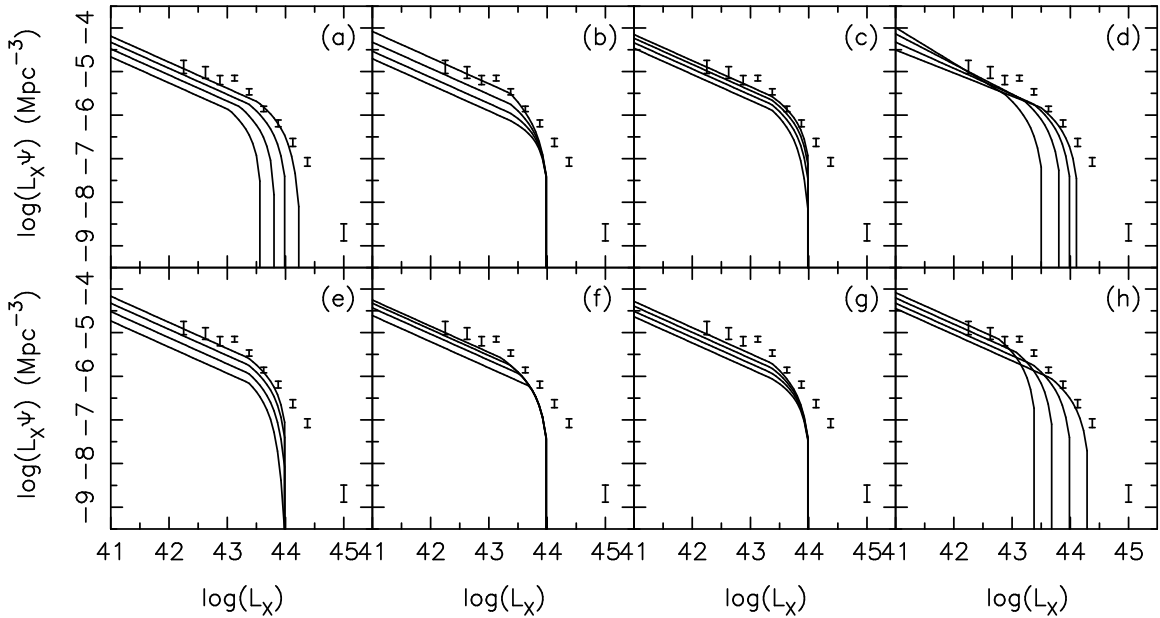}
\caption{The X-ray LF of tidal disruptions calculated using equation (\ref{eq:psi}) assuming a bolometric correction $\omega=45$ and a light curve that decays as a power-law $\lambda=\frac{5}{3}$.  In each panel, a parameter is varied to illustrate dependence: (a) $m_*=(0.1,0.333,{\it 1.0},3.0)M_\odot$; (b)  $\alpha=(-1,-1.15,{\it -1.3},-1.45)$;   (c)  $k=(0.70,0.75,{\it 0.8},0.85)$; (d) $\lambda=1.222,1.444,{\it 1.666},1.888$; (e) $M_{B\star}=(-18,-19,{\it -20},-21)+5\log_{10}h$;  (f) $M_{\rm min}=(10^3,{\it 10^4},10^5,10^6)M_\odot$;  (g) $a=(4.0,4.333,{\it 4.666},5.0)$; and (h) $\xi=0.5,{\it 1},2,4$, where the fiducial value has been italicized. We also show the low-redshift ($0.015<z<0.2$) soft X-ray ($0.5-2\textrm{ keV}$) LF data from the {\it Chandra} and {\it XMM-Newton} survey of \citet{Hasinger:05}. A Hubble constant of $H_0=70\textrm{ km s}^{-1}\textrm{ Mpc}^{-1}$ is assumed throughout.\label{fig:luminosity_function}}
\end{figure*}

In Figure \ref{fig:luminosity_function}, we plot the resulting LF for our fiducial model with a light curve that decays as a power-law $L\propto t^{-\lambda}$ with $\lambda=\frac{5}{3}$.  The model is based on a galaxy LF with amplitude $\Phi_0=0.02\ h^{-3}\textrm{ Mpc}^{-3}$ and absolute blue magnitude of an $L_\star$ galaxy of $M_{B\star}=-20.0+5\log_{10}h$ \citep{Blanton:01}, where we assume $h=0.7$.  We vary the various parameters, including $\lambda$ in Figure \ref{fig:luminosity_function}\emph{d}, to explore the sensitivity of the X-ray AGN LF to our choice of the fiducial model.  In the low-luminosity regime $L_{\rm X}\ll L_{\rm peak}\equiv \omega^{-1} L_{\rm bol}[m_\star,t_{\rm peak}(m_\star)]$, the LF is a power-law $\Phi(L_{\rm X})\propto L_{\rm X}^{-1/\lambda-1}$.  At $L_{\rm X}\gtrsim L_{\rm peak}\sim 10^{44} (\omega/45)^{-1} m_\star^{1/3} \textrm{ ergs s}^{-1}$, the LF drops precipitously. 

Below $L_{\rm X}\sim 10^{44}\textrm{ ergs s}^{-1}$, the calculated X-ray AGN LF slightly overestimates but is roughly compatible with the soft X-ray AGN LF of \citet{Hasinger:05} (shown in the figure), and is a factor of a few lower than the hard X-ray AGN LF \citep{Ueda:03,Barger:05b}.  The discrepancy between the soft and hard X-ray LF could be attributed to an incompleteness of the soft X-ray LF resulting from obscuration.

In models with power-law decaying light curves, the low-luminosity slope of the AGN LF is determined by the power-law of the decay, rather than the shape of the black hole mass function (Fig.~\ref{fig:luminosity_function}\emph{d}).  The amplitude of the LF, however, is sensitive to the black hole mass function.  The steep drop in the LF around $10^{44}\textrm{ ergs s}^{-1}$ is associated with the peak luminosity produced in a tidal disruption event, which corresponds to the peak mass inflow rate from the debris that is beginning to circularize around the black hole (Fig.~\ref{fig:luminosity_function}\emph{a,h}).  

In the ``while the supplies last'' model, where the accretion luminosity is a constant and equal to $\ell L_{\rm Edd}$, the LF reads
\beq
\Psi(L_{\rm X})\approx  \frac{ \epsilon }{\ell}\frac{\sigma_{\rm T}  c}{4\pi G m_{\rm p} }\frac{m_\star}{L_{\rm X}}\Gamma(M_1)  \Psi(M_1) ,
\eeq
for $M_{\rm min}<M_1<M_{\rm max}$, where $M_1=\omega\sigma_{\rm T}L_{\rm X}/4\pi\ell G m_{\rm p} c$.  In the low luminosity regime, the LF is a power law $\Psi(L_{\rm X})\propto L_{\rm X}^{-3+7/2a+(\alpha+1)k}\ \ell^{1-7/2a-(\alpha+1)k}$,
which assuming a constant $\ell$ is steeper, $L_{\rm X}\Psi(L_{\rm X})\propto L_{\rm X}^{-1.5}$, than the  observed relation ($\mu>-1$), although the two LFs have comparable amplitudes at $L_{\rm X}\sim 10^{43}\textrm{ ergs s}^{-1}$.  The discrepancy may reflect a luminosity-dependent Eddington ratio $\ell$.  The dependence of $\ell$ on luminosity or black hole mass is expected if the accretion rate is supply limited, rather than radiatively limited.   For example, to reproduce $\Psi(L_{\rm X})\propto L_{\rm X}^{-2}$, we must have $\ell \propto L_{\rm X}$.  At present it is not clear how such dependence should arise.

%\section{Results}
%\label{sec:results}

\section{Discussion}
\label{sec:discussion}

While many theoretical uncertainties frustrate an accurate estimate of the contribution of activity associated with stellar tidal disruption to the X-ray LF of active galaxies, it seems that the contribution can be significant at low luminosities $L_{\rm X}\lesssim 10^{44}\textrm{ ergs s}^{-1}$.  The tidal disruption activity makes a negligible contribution at higher luminosities (power-law profile of the measured soft X-ray LF extends to $L_{\rm X}\sim 10^{47.5}\textrm{ ergs s}^{-1}$ at redshifts $z\sim 1-3$).  MBHs with masses above $\sim 10^7 M_\odot$ grow by accreting non-stellar material from the interstellar medium.
%\footnote{In an alternative interesting model, quasar accretion disks capture stars and the black hole grows by accreting stellar debris \citep[and references therein]{MiraldaEscude:05}.}  

What is the origin of the featureless knee in the LF at $L_{\rm X}\sim (10^{43}-10^{44})\textrm{ ergs s}^{-1}$?  The location of the knee coincides with Eddington-limited accretion onto black holes with masses $M_{\rm bh}\sim 10^{6-7}M_\odot$.\footnote{The maximum accretion rate might exceed the Eddington limit by a factor of a few (e.g., \citealt{Begelman:02}).}  Given the tidal disruption rate assumed here (eq.~\ref{eq:n_dot}), black holes with masses $M_{\rm bh}\lesssim 2\times10^6(f/0.25)M_\odot$, where $f$ is the fraction of the stellar mass that ultimately gets accreted onto the MBH, could have grown to their present size by accreting stellar debris over a Hubble time, although undoubtedly, non-stellar accretion must have played a role.  This critical mass is compatible with the location of the knee in the LF assuming near-Eddington accretion.  

Black holes with masses $\sim 10^7M_\odot$, in turn, roughly lie at the demarcation line separating the dominance of late-type and early-type galaxies.  If major mergers are responsible for diverting gas into accretion onto black holes in early type galaxies, then the absence of merging activity in late-type galaxies leaves stellar tidal disruption as competitive supplier of material to the MBH.  This may explain why the steep power-law behavior of the quasar LF fails to extend to lower luminosities.

The observed X-ray LF of AGN evolves with redshift.  The number density of low-luminosity AGN increases by a factor of $\sim (10,4)$ at $L_{\rm X}\sim (10^{42},10^{43})\textrm{ ergs s}^{-1}$ from $z\sim 0$ to $z\sim 0.7$ in the soft X-rays ($0.5-2\textrm{ keV}$; \citealt{Hasinger:05}), while the respective increase is only a factor $\sim 2-3$ in the hard X-rays ($2-8\textrm{ keV}$; \citealt{Barger:05b}). The location of the knee shifts from $L_{\rm X}\sim10^{43}\textrm{ ergs s}^{-1}$ to $L_{\rm X}\sim10^{44}\textrm{ ergs s}^{-1}$ from $z\sim 0$ to $z\sim 1$ in the soft and hard band alike, and the break becomes more prominent at higher redshift.  These trends require explanation.  

The number density of tidal-disruption powered sources could increase with redshift because of one or more of the following reasons: 1. On average, MBHs are smaller at larger redshift, implying a larger tidal disruption rate or, speculatively, a higher radiative efficiency in the soft X-ray band;\footnote{Note that for black holes smaller than $M_{\rm bh}\lesssim 10^6M_\odot$, thermal emission from an Eddington-limited disk enters the soft X-ray band, and the soft X-ray luminosity is thus anticorrelated with mass.  It is also possible that transition from high to low-state accretion is sensitive to $M_{\rm bh}$.}  2. Galactic nuclei were denser at larger redshift (or, equivalently, stellar velocity dispersions were higher at fixed MBH mass), also implying a larger tidal disruption rate;
%\footnote{Two body relaxation can induce an expanion in the nucleus \citep[and references therein]{Murphy:91}.} 
and 3. Tidally disrupted stars were more massive at larger redshift, implying brighter events.\footnote{The remaining possibility, an increase of the number of black holes in the centers of galaxies with redshift, seems implausible.}  Any one or all of these factor may contribute to the observed trend at low luminosities, and could also be responsible for the apparent brightening of the location of the knee of the LF.  

In particular, if MBH grew gradually in pre-existing stellar density cusps with fixed velocity dispersions, then the tidal disruption rate would be inversely proportional to the black hole mass (eq.~\ref{eq:n_dot}) and would increase with redshift.  This could explain the increase in the number density of low-luminosity AGN from the nearby universe to $z\sim 0.7$.

The relative contribution of stellar tidal disruptions to the X-ray AGN LF can be studied by examining long-term X-ray variability of the sources.  Currently, long-term monitoring data have been analyzed for only a handful of sources (e.g., \citealt{Markowitz:04}).  Theoretical uncertainty notwithstanding, the light curves of tidal disruption events should be characterized by a fast rise and slow decay, but the variability may or may not be detected as an outburst.  The frequency of hard X-ray outbursts in the {\it ROSAT} All-Sky Survey was studied by \citet{Donley:02}.  Their estimate, $\sim10^{-5}\textrm{ yr}^{-1}$ per galaxy, is significantly below the tidal disruption rate adopted by us. The rates can be reconciled if the majority of tidal disruptions do not exhibit outburst characteristics.  In that case, a measurable  asymmetry in the autocorrelation function will still be expected.

\acknowledgements
 
We thank Avishay Gal-Yam, Eran Ofek, Sterl Phinney, and Ohad Shemmer for inspiring discussions. M.~M.\ acknowledges support by NASA through the  Hubble Fellowship grant HST-HF-01188.01-A awarded by the Space Telescope Science Institute, which is operated by the Association of Universities for Research in Astronomy, Inc., for NASA under contract NAS5-26555.  D.~M.\ acknowledges support by NSF grants AST-0420920 and AST-0437519 and NASA grant NNG04GJ48G.

\end{document}